\documentclass[3p,times]{elsarticle}

\usepackage{amsmath}
\newtheorem{prop}{Conjecture}

\usepackage{amssymb}

\usepackage[figuresright]{rotating}

\begin{document}

\begin{frontmatter}




\title{Products of Bessel functions and associated polynomials}


\author[Enea]{G. Dattoli \corref{cor}}
\ead{giuseppe.dattoli@enea.it}

\author[Enea]{E. Di Palma }
\ead{emanuele.dipalma@enea.it}

\author[Enea]{E. Sabia}
\ead{elio.sabia@enea.it}

\author[Enea,Unipa]{S. Licciardi }
\ead{silvia.licciardi@dmi.unict.it}

\address[Enea]{ENEA - Centro Ricerche Frascati, Via Enrico Fermi 45, 00044, Frascati, Rome, Italy}
\cortext[cor]{Corresponding author}
\address[Unipa]{University of Palermo, Department of Mathematics, Via Archirafi, 34, 90123 Palermo, Italy}

\begin{abstract}
Symbolic methods of umbral nature are exploited to derive series expansion for the products of Bessel functions. It is shown that the
product of two cylindrical Bessel functions can be written in terms of Jacobi polynomials. The procedure is extended to products of an arbitrary number of functions and the link with previous researchers is discussed. We show that the technique we propose and the use of the Ramanujan  master theorem allow the derivation of integrals of practical interest.
\end{abstract}

\begin{keyword}
Bessel Functions, Hermite Polynomials, Umbral Calculus
\end{keyword}

\end{frontmatter}


\section{Introduction}
\addcontentsline{toc}{chapter}{Introduzione}{}
\markboth{\textsc{Introduzione}}{}

In a recent paper Moll and Vignat \cite{Vignat} have considered the series expansion of powers of the modified Bessel function $(BF)$ of first kind.
 They found that the expansion procedure involves a family of polynomials introduced in \cite{Bender} and,  upon extending the relevant properties, the authors obtained a link with the umbral formalism of ref. \cite{Cholewinski}. \\
 In a previous paper \cite{Brychkov} an analogous problem has been addressed by Brychkov who provided a formalism for the derivation of
 products of special functions in general and of BF in particular.

 We reconsider here the problem addressed in refs \cite{Vignat, Brychkov}, within the framework of the formalism (also of umbral nature) developed in \cite{DBabusci}, which will be reviewed in this
  introduction. We will prove that it is naturally suited  to obtain the power series of the product of two BF's and in the forthcoming sections we will discuss the extension to any arbitrary number.\\

One of the main results of \cite{DBabusci} has been the conclusion that the \textit{0-th} order cylindrical BF is the Umbral \textit{(U-)} image of a Gaussian function. By setting indeed that
\begin{equation}
J_{0}(x)=e^{-\tilde{c} \left(\frac{x}{2} \right)^{2} }\varphi_{0}
\label{eqn:form1}
\end{equation}
where the \textit{U-} operator $\tilde c$ is defined in such a way that
\begin{equation}
\tilde c^{\nu}\varphi_{0}=\dfrac{1}{\varGamma(\nu+1)}
\end{equation}
with $\nu$ being not necessarily a positive real integer and, using an expression partially borrowed from field theory, $\varphi_{0}$ will be said the \textit{U-"vacuum"}.
From the previous identities we recover  the expansion
\begin{equation}
e^{-\tilde{c} \left(\frac{x}{2} \right)^{2} }\varphi_{0}=\sum_{r=0}^{\infty}\dfrac{(-1)^{r}}{r!}\left(\dfrac{x}{2} \right)^{2r}\tilde c^{r}\varphi_{0}= \sum_{r=0}^{\infty}\dfrac{(-1)^{r}}{(r!)^{2}}\left(\dfrac{x}{2} \right)^{2r}
\end{equation}
which is the ordinary series defining the \textit{0-th} order cylindrical Bessel \cite{Andrews}.\\
According to eq.(\ref{eqn:form1}) a \textit{0-th} order cylindrical BF is essentially a Gaussian and, provided that the operator $\tilde c$ can be treated as an ordinary algebraic quantity, they can be handled by taking advantage from the elementary properties of exponential functions.
Even though the notion of vacuum is extremely doubtful from the mathematical point of view it will be used here as a computational tool, a more appropriate definition is provided in the concluding section. \\
Let us now consider the product
\begin{equation}
f(x;a,b)= J_{0}(ax)J_{0}(bx)
\end{equation}
which can be formally written as the product of two Gaussians , namely \footnote{Even though not explicitly stated, it is evident that in the present formalism we have \begin{equation}
\left[J_{0}(x) \right]^{2}= e^{-\tilde{c} \left(\frac{x}{2} \right)^{2} }e^{-\tilde{c} \left(\frac{x}{2} \right)^{2} }\varphi_{0} \nonumber
\end{equation}}
\begin{equation}
f(x;a,b)=e^{-(a^{2}\tilde c_{1}+b^{2}\tilde c_{2})\left(\frac{x}{2} \right)^{2} }\varphi_{0}^{(1)}\varphi_{0}^{(2)}
\end{equation}
where $\varphi_{0}^{(\alpha)}$ are the \textit{U-vacua} on which the operators $\tilde c^{\alpha}$ act. The series expansion of the exponential and the use of the previously outlined rules yield
\begin{equation}\begin{split}
& f(x;a,b)=\sum_{r=0}^{\infty}\dfrac{(-1)^{r}}{r!}l_{r}(a^{2},b^{2})\left(\frac{x}{2} \right)^{2r}\\
& l_{r}(a,b)=r!\sum_{s=0}^{r}\dfrac{a^{(r-s)}b^{s}}{(s!)^{2}\left[(r-s)! \right]^{2} }
\end{split}\end{equation}
In case of $a=b$ the expression for $f(x;a,b)$ is equivalent to that reported in ref \cite{Vignat}
\begin{equation}
f(x;a,a)=\sum_{r=0}^{\infty}\dfrac{(-1)^{r}}{[r!]^{2}}B_{r}(2)\left(\dfrac{ax}{2} \right)^{2r}
\end{equation}
where $B_{r}(2)$ is calculated with a recursive formula
\begin{equation}
B_{n}(r)=\sum_{s=0}^{n}\binom {n}{s}\dfrac{n!}{s!(n-s)!}B_{s}(r-1) \qquad \mbox{with} \;\;\; B_{0}(r)=0;\;B_{n}(0)=\delta_{n,0}
\end{equation}
Leaving for the moment unspecified the nature of the polynomials $l_{r}(a,b)$, we note that the function $f(x;a,b)$ can be cast in the \textit{U-}form
\begin{equation}\begin{split}\label{f(xab)}
& f(x;a,b)=e^{-\tilde l \left(\frac{x}{2} \right)^{2} }\Phi_{0}\\
& \tilde l^{\nu}\Phi_{0}=l_{\nu}(a^{2},b^{2})
\end{split}\end{equation}
The action of the operator $\tilde l$ on the corresponding \textit{U-}vacuum holds for any real (positive/negative) or complex value of the  exponent $\nu$.
We have concluded that  the product of two cylindrical Bessel is the \textit{U-}equivalent of a $BF$ and thus the umbra of a Gaussian. Such a conclusion turns particularly useful if we are interested in the evaluation of the integrals of the function $f(x;a,b)$, a straightforward use of the so far developed procedure yields
\begin{equation}\begin{split}\label{Ramanj}
& \int_{-\infty}^{+\infty}f(x;a,b)dx=\int_{-\infty}^{+\infty}e^{-\tilde l \left(\frac{x}{2} \right)^{2} }dx\Phi_{0}=2\sqrt{\pi}\tilde l^{-\frac{1}{2}}\Phi_{0}=2\sqrt{\pi}l_{-\frac{1}{2}}(a^{2},b^{2}),\qquad \mid a\mid>\mid b\mid  \\
& l_{-\frac{1}{2}}(a^{2},b^{2})=\Gamma \left(\dfrac{1}{2} \right)\sum_{s=0}^{\infty}\dfrac{a^{-2\left(\frac{1}{2}+s \right) }b^{2s}}{(s!)^{2}\Gamma \left(\dfrac{1}{2}-s \right)^{2}}=\dfrac{1}{\sqrt{\pi} \mid a\mid}K\left(\dfrac{b}{a} \right), \\
& K(k)={}_{2}F_{1}\left(\dfrac{1}{2},\dfrac{1}{2};1;k^{2} \right)= \sum_{s=0}^{\infty}\left[\dfrac{(2s)!}{2^{2s}(s!)^{2}} \right]^{2}k^{2s}
\end{split}\end{equation}
This result can however be viewed as an application of the Ramanujan master theorem \cite{Ramanujan}, it has, indeed, been derived by treating the \textit{U-}operator $\tilde l$ as an ordinary constant and then by applying the rules of the Gaussian integrals. The correctness of the result has then been checked numerically.\\

We have left open the question on the nature of the polynomials $l_{r}(a,b)$, although we will discuss more deeply this point in the forthcoming sections, here we note that they can be viewed as a particular case of the Jacoby polynomials \cite{Andrews}, as it can be inferred from the identity \cite{BabDatt}:
\begin{equation}\begin{split}
& l_{r}\left(\dfrac{x-1}{2},\dfrac{x+1}{2} \right)=\dfrac{1}{r!}P_{r}^{(0,0)}(x)  \\
& P_{n}^{(\alpha ,\beta)}(x)=\sum_{s=0}^{n}\binom {n+\alpha}{s}\binom {n+\beta}{n-s}\left( \dfrac {x-1}{2}\right) ^{n-s}\left( \dfrac {x+1}{2}\right) ^{s}
\end{split}\end{equation}
Furthermore, since in \textit{U-}form the cylindrical Bessel functions of \textit{n-th} order read \cite{Andrews}
\begin{equation}\label{cylBess}
J_{\nu}(x)=\left(\dfrac{x}{2} \right)^{\nu}\tilde c^{\nu}e^{-\tilde c\left(\frac{x}{2} \right)^{2}}\varphi_{0}
\end{equation}
we obtain the following general expression for the product of two cylindrical Bessel functions of order $\nu ,\mu$ respectively
\begin{equation}\begin{split}
& f_{\nu ,\mu}(x;a,b)=J_{\nu}(ax)J_{\mu}(bx)=\left(\dfrac{x}{2} \right)^{\nu+\mu}(a\tilde c_{1})^{\nu}(b\tilde c_{2})^{\mu}e^{-(a^{2}\tilde c_{1}+b^{2}\tilde c_{2})\left(\frac{x}{2} \right)^{2} }\varphi_{0}^{(1)}\varphi_{0}^{(2)}=  \\
& =\left(\dfrac{x}{2} \right)^{\nu+\mu}a^\nu b^\mu\sum_{r=0}^{\infty}\dfrac{(-1)^{r}}{r!}l_{r}^{(\nu ,\mu)}(a^{2},b^{2})\left( \dfrac{x}{2}\right)^{2r}; \\
& l_{r}^{(\nu ,\mu)}(a,b)=r!\sum_{s=0}^{r}\dfrac{a^{(r-s)}b^{s}}{\Gamma (\mu+s+1)\Gamma(\nu+r-s+1)s!(r-s)!}
\end{split}\end{equation}
We have so far provided a first idea of how the $U-$formalism of ref. \cite{Andrews} works and how it can be exploited to study the properties of products of (cylindrical) Bessel functions,  in the forthcoming section we will take advantage from its simplicity to extend the method to arbitrary products.

\section{Products of Bessel functions}

According to the tools outlined in the previous section,  the product of three \textit{0-th} order Bessel functions, can be written as
\begin{equation}
f(x;a_{1},a_{2},a_{3})=e^{-(a_{1}^{2}\tilde c_{1}+a_{2}^{2}\tilde c_{2}+a_{3}^{2}\tilde c_{3})\left(\frac{x}{2} \right)^{2} }\varphi_{0}^{(1)}\varphi_{0}^{(2)}\varphi_{0}^{(3)}
\end{equation}
Or, in explicit form
\begin{equation}\begin{split}\label{1.2}
& f(x;a_{1},a_{2},a_{3})=\sum_{r=0}^{\infty}\dfrac{(-1)^{r}}{r!}l_{r}(a_{1}^{2},a_{2}^{2},a_{3}^{2})\left(\dfrac{x}{2} \right)^{2r}\\
& l_{r}(x_{1},x_{2},x_{3})=r!\sum_{s=0}^{r}\dfrac{x_{3}^{(r-s)}}{(s!)[(r-s)!]^{2}}l_{s}(x_{1},x_{2})
\end{split}\end{equation}
It is evident that the extension to the case of $n$ $BF$ writes as in the first of eqs.(\ref{1.2}) with
\begin{equation}
l_{r}(x_{1},\dots ,x_{n})=r!\sum_{s=0}^{r}\dfrac{x_{n}^{(r-s)}}{(s!)[(r-s)!]^{2}}l_{s}(x_{1},\dots,x_{n-1})
\end{equation}
in ref. \cite{Vignat} all the $\alpha$ parameters  (actually the variables of the $l$ polynomials) are assumed to be $1$.\\

From a formal point of view  the use of the multinomial expansion allows to define the previous family of polynomials as
\begin{equation}
l_{r}(x_{1},\dots,x_{n})=(x_{1}\tilde c_{1}+\dots +x_{n}\tilde c_{n})^{r}\varphi_{0}^{(1)}\dots \varphi_{0}^{(n)}
\end{equation}
and the use of the multinomial expansion yields
\begin{equation}
l_{r}(x_{1},\dots ,x_{n})=\sum_{k_{1}+\dots +k_{n}=r}\binom {r}{k_{1} \;\dots\; \dots \;\dots\; k_{r}}\dfrac{x_{1}^{k_{1}}}{(k_{1}!)^{2}} \dots \dfrac{x_{n}^{k_{n}}}{(k_{n}!)^{2}}
\end{equation}
Going back to the two variable case it is easy to check that they satisfy the differential equation
\begin{equation}
\partial _{x_{1}}x_{1}\partial _{x_{1}}l_{r}(x_{1},x_{2})=\partial _{x_{2}}x_{2}\partial _{x_{2}}l_{r}(x_{1},x_{2})=rl_{r-1}(x_{1},x_{2})
\end{equation}
With $\partial _{x}x\partial _{x}$ being the so called Laguerre derivative \cite{G.Dattoli}. The Laguerre polynomials can indeed cast in the form of eq.(\ref{cylBess})
\begin{equation}\label{Laguerre}
L_{n}(x,y)=(y-\tilde c_{1}x)^{n}\varphi_{0}^{(1)}
\end{equation}
To obtain the extension to the product of arbitrary cylindrical Bessel, it will be sufficient to replace in the previous equations the function $l_{r}(a_{1}^{2}, \dots , a_{n}^{2})$ with $l_{r}^{(\nu_{1}, \dots ,\nu_{n})}(a_{1}^{2}, \dots ,a_{n}^{2})$
\begin{equation}\begin{split}
& l_{r}^{(\nu_{1}, \dots ,\nu_{n})}(x_{1}, \dots ,x_{n})=\tilde c_{1}^{\nu_{1}} \dots \tilde c_{n}^{\nu_{n}}(x_{1}\tilde c_{1}+\dots +x_{n}\tilde c_{n})^{r}\varphi_{0}^{(1)}\dots \varphi_{0}^{(n)}= \\
& =\sum_{k_{1}+\dots +k_{n}=r}\binom {r}{k_{1} \;\dots \;\dots \;\dots\; k_{r}}\dfrac{x_{1}^{k_{1}}}{k_{1}!\varGamma (\nu_{1}+k_{1}+1)} \dots \dfrac{x_{n}^{k_{n}}}{k_{n}!\varGamma (\nu_{n}+k_{n}+1)}
\end{split}\end{equation}
Thus getting an expression closely similar to that derived by Brychkov in \cite{Brychkov}
\begin{equation}
\prod_{s=1}^{n}J_{\nu_{s}}(a_{s}x)=\left(\dfrac{x}{2} \right)^{\sum_{s=1}^{n}\nu_{s}}\left(\prod_{k=1}^{n}a_{k}^{\nu_{k}} \right)\sum_{r=0}^{\infty}\dfrac{(-1)^{r}}{r!}l_{r}^{(\nu_{1},\dots , \nu_{n})}\left( a_{1}^{2},\dots ,a_{n}^{2}\right) \left(\dfrac{x}{2} \right)^{2r}
\end{equation}
In the case of modified $BF$ of first kind the procedure is the same, the function can be formally expressed as a quadratic exponential and we can recover the results of ref. \cite{Vignat}, by noting that the functions actually used in that paper the $BF$  are given by
\begin{equation}
\tilde I_{\nu}(x)=\sum_{r=0}^{\infty}\dfrac{\Gamma (\nu +1)}{r!\Gamma (r+\nu +1)}\left(\dfrac{x}{2} \right)^{2r}=\Gamma (\nu+1)\tilde c^{\nu}e^{\tilde c\left(\frac{x}{2} \right)^{2}}\varphi_{0}
\end{equation}
According to our formalism the relevant \textit{k-th} power reads
\begin{equation}
(\tilde I_{\nu}(x))^{k}=\Gamma (\nu+1)^{k}\sum_{r=0}^{\infty}\dfrac{1}{r!}l_{r}^{(\nu,\dots , \nu)}(1,\dots , 1)\left(\dfrac{x}{2} \right)^{2r}
\end{equation}
The polynomials defined in \cite{Vignat} are expressible in terms of our $l_{r}^{(\nu_{1},\dots , \nu_{n})}(x_{1},\dots , x_{n})$ as
\begin{equation}\begin{split}
& B_{r}^{(\nu)}(k)=\Gamma(\nu+1)^{k-1}\Gamma(r+\nu+1)l_{r}^{\{\nu\}}(k)\\
& l_{r}^{(\nu,\dots , \nu)}(1,\dots , 1)=l_{r}^{\{\nu\}}(k) \\
& l_{r+1}^{\{\nu\}}(k)=\sum_{j=1}^{k}l_{r}^{\{\nu+1_{j}\}}(k)\\
& \{\nu+1_{j},k\}=(\nu,\dots , \nu+1, \dots ,\nu)
\end{split}\end{equation}
The nature of the polynomials $l_{r}^{\{\nu\}}(k)$ will be further discussed in the forthcoming concluding section.

\section{Final Comments and Applications}

In the previous sections we have used umbral concepts without an appropriate justification.
The operators $\tilde c$ have been assumed to act on a naively defined "vacuum" $\varphi_{0}$
introduced by borrowing the terminology and the formalism from quantum field theory.
Notwithstanding the results we have obtained in the present note and in precedent papers as well have been shown to be fairly efficient computational tools.\\
We will frame the previous definitions within a more rigorous context by the use of Borel-Laplace transform (\textit{BLT}) method. \\
Given a function  $f_B(x)$ we define the relevant BLT \cite{Ehrenpreis}
\begin{equation}
f_B(x)=\int_0^{\infty} e^{-t}f(tx) dt
\end{equation}
Such a definition can be further elaborated by using the operational identity \cite{Ramanujan}
\begin{equation}
f(tx)=t^{x \partial_x}f(x)
\end{equation}
which allows to write
\begin{eqnarray}
f_B(x)=\hat{B} [f(x)] \nonumber \\ \\
 \hat{B}=\int_0^{\infty} e^{-t} t^{x \partial_x} dt=\Gamma(x \partial_x +1)   \nonumber
\end{eqnarray}
The BLT  can be generalized by introducing its $\alpha-th$  order counterpart; by defining indeed the operator
\begin{equation}
 \hat{B}^{\alpha}=\int_0^{\infty} e^{-t} t^{\alpha x \partial_x} dt=\Gamma(\alpha x \partial_x +1)
\end{equation}
we set
\begin{equation}
f_B^{(\alpha)}=\hat{B}^{(\alpha)}[f(x)]=\int_0^{\infty} e^{-t} f(t^{\alpha} x)dt
\end{equation}
In the case of the $\alpha=\frac{1}{2}$,   BLT applied to the \textit{0-th} order Bessel yields
\begin{equation}
\hat{B}^{(\frac{1}{2})}[J_0(x)]=\Gamma\left(\frac{1}{2}x \partial_x+1\right) \sum_{r=0}^{\infty} \frac{(-1)^r}{(r!)^2}
\left(\frac{x}{2}\right)^{2r}=\sum_{r=0}^{\infty}\frac{(-1)^r}{r!} \left(\frac{x}{2} \right)^{2r}
\end{equation}
By assuming that the $\alpha$ order BLT admits an inverse, namely that
\begin{eqnarray}
\left(\hat{B}^{(\alpha)} \right)^{-1}  \hat{B}^{(\alpha)}&=& \hat{1}  \nonumber \\ \\
 \left(\hat{B}^{(\alpha)} \right)^{-1} &=&\frac{1}{\Gamma(\alpha x \partial_x+1)}   \nonumber
 \label{eqn:fora}
\end{eqnarray}
we can also state that
\begin{equation}
\left[ \hat{B}^{\left( \frac{1}{2}\right)} \right]^{-1} \left[ e^{-\left(\frac{x}{2}\right)^2}\right]=J_0(x)
\end{equation}

The extension of eq.(\ref{eqn:fora}) to negative  $\alpha$ yields

\begin{equation}
\hat{B}^{(-\alpha)}=\Gamma(-\alpha x \partial_x+1)=\frac{1}{\Gamma(\alpha x \partial_x)}\frac{\pi}{sin(\alpha \pi x \partial_x)}
\end{equation}

and it is worth stressing that

\begin{equation}
\hat{B}^{(-\alpha)}\neq \left[\hat{B}^{(\alpha)}  \right]^{-1}
\end{equation}
After the previous remarks we can state the following proposition: \\
\begin{prop}
Given the function  having the integral
\begin{equation}
\int_{-\infty}^{\infty} f(x)dx=k
\end{equation}
then
\begin{equation}
\int_{-\infty}^{\infty} \hat{B}^{(\alpha)} \left[f(x)\right]dx=k \Gamma(1-\alpha)
\end{equation}
\end{prop}
which can be viewed as a restatement of the Ramanujan Master theorem \cite{Ramanujan}. \\ \\
Postponing the proof of our conjecture to a forthcoming dedicated note, we provide here an example, regarding the particular case of Bessel functions.\\
According to the previous definitions we obtain
\begin{eqnarray}
\int_{-\infty}^{\infty} e^{-\left(\frac{x}{2}\right)^2} dx&=& \int_0^{\infty} dt e^{-t} \int_{-\infty}^{\infty} J_0( \sqrt{t}x )dx=I_{J_0}\Gamma\left(\frac{1}{2}\right) \nonumber \\ \label{eqnM}\\
I_{J_0}&=& \int_{-\infty}^{\infty} J_0(x)dx   \nonumber
\end{eqnarray}
assuming $I_{J_0}$  unknown, we find from eq. (\ref{eqnM})
\begin{equation}
I_{J_0}= \left[ \Gamma\left(\frac{1}{2} \right) \right]^{-1} \int_{-\infty}^{\infty} e^{-\left( \frac{x}{2} \right)^2} dx=2
\end{equation}
Analogous statements hold for the "transposition" of other properties of the Gaussian function (like those under derivative) to the Bessel function. \\
The previous remarks suggest that integral transform of Laplace type can be exploited as the appropriate environment to justify the umbral  formalism developed so far and which will be further extended in this section.\\
We go therefore back to the polynomials  $l_{r}^{(\nu_{1},\dots , \nu_{n})}(x_{1},\dots , x_{n})$, which can also be associated with multidimensional Jacobi like polynomials.\\

We observe in particular that the relevant generating function is expressible in terms of product of Bessel like functions, namely
\begin{equation}\begin{split}
& \sum_{r=0}^{\infty}\dfrac{(-t)^{r}}{r!}l_{r}^{(\nu_{1},\dots , \nu_{n})}(x_{1},\dots , x_{n})=\prod_{j=1}^{n}C_{\nu_{j}}(tx_{j})\\
& C_{\nu}(x)=\sum_{r=0}^{\infty}\dfrac{(-x)^{r}}{r!\Gamma(\nu+r+1)}
\end{split}\end{equation}
where $C_{\nu}(x)$ denotes the Bessel Tricomi function \cite{BabDatt} of order $\nu$ \footnote{The link between Tricomi and Bessel functions is provided by \begin{equation}
C_{\nu}(x)=\left(x \right)^{-\frac{\nu}{2}}J_{\nu}(2\sqrt{x})
\nonumber \end{equation}  }.\\

The polynomials $l_{r}^{\{\nu\}}(k)$ are something else
\begin{equation}
l_{r}^{\{\nu\}}(k)=\tilde c_{1}^{\nu} \dots \tilde c_{k}^{\nu}(\tilde c_{1}+\dots +\tilde c_{k})^{r}\varphi_{0}^{(1)}\dots \varphi_{0}^{(k)}
\end{equation}
We find that
\begin{equation}\begin{split}
& l_{r}^{\{\nu\}}(k+1)=\tilde c_{k+1}^{\nu}\tilde c_{1}^{\nu} \dots \tilde c_{k}^{\nu}(\tilde c_{1}+\dots +\tilde c_{k}+\tilde c_{k+1})^{r}\varphi_{0}^{(1)}\dots \varphi_{0}^{(k)}\varphi_{0}^{(k+1)}=\\
& =\tilde c_{k+1}^{\nu}\sum_{j=0}^{r}\binom {r}{j}\tilde c_{k+1}^{r-j}l_{j}^{\left\lbrace \nu\right\rbrace }(k)\varphi_{0}^{(k+1)}=\sum_{j=0}^{r}\binom {r}{j}\dfrac{1}{\Gamma (r-j+\nu+1)}l_{j}^{\left\lbrace \nu\right\rbrace }(k)
\end{split}\end{equation}
The various identities reported in \cite{Vignat} follow from the above equation, which can be generalized in various ways, as e.g.
\begin{equation}\begin{split}
& l_{r}^{\{\nu\}}(k+s)=\tilde c_{k+1}^{\nu}\dots\tilde c_{k+s}^{\nu}\tilde c_{1}^{\nu}\dots \tilde c_{k}^{\nu}(\tilde c_{1}+\dots +\tilde c_{k}+\tilde c_{k+1}+\dots +\\
& +\dots+\tilde c_{k+s})^{r}\varphi_{0}^{(1)}\dots \varphi_{0}^{(k)}\varphi_{0}^{(k+1)}\dots \varphi_{0}^{(k+s)}=\\
& =\sum_{j=0}^{r}\binom {r}{j}l_{r-j}^{\{\nu\}}(s)l_{j}^{\{\nu\}}(k)
\end{split}\end{equation}
The use of the $U-$formalism has been proven to be a powerful tool allowing a very quick understanding of the various technicalities underlying the handling of products or powers of Bessel functions.\\

We have noted in the introductory section that the use of straightforward algebraic manipulations allows the derivation of an expression yielding the integral of the product of two cylindrical Bessel functions. We have checked that the extension to the products of three or more is anyway feasible.
Regarding the case of an integral of the product of three Bessel functions we find
\begin{equation}\begin{split}
& \int_{-\infty}^{+\infty}f(x;a_{1},a_{2},a_{3})dx=2\sqrt{\pi}l_{-\frac{1}{2}}(a_{1}^2,a_{2}^2,a_{3}^2), \qquad \mid a_{3}\mid > \mid a_{2}\mid > \mid a_{1}\mid \\
& l_{-\frac{1}{2}}(a_{1},a_{2},a_{3})=\Gamma\left( \dfrac{1}{2}\right) \sum_{s=0}^{\infty}\dfrac{a_{3}^{-\left(\frac{1}{2}+s \right) }}{(s!)\Gamma\left(\dfrac{1}{2}-s \right)^{2} }l_{s}(a_{1},a_{2})
\end{split}\end{equation}
In eq.(\ref{Ramanj}) we have recognized that the series defining $l_{-\frac{1}{2}}(a,b)$ can be recognized as that defining a quarter period elliptic integral, in this case we obtain
\begin{equation}\begin{split}
& l_{-\frac{1}{2}}(a_{1},a_{2},a_{3})=\dfrac{1}{\sqrt{\pi \mid a_{3}\mid}} F(a_{1},a_{2},a_{3})\\
& F(a_{1},a_{2},a_{3})=\sum_{s=0}^{\infty}\left[\dfrac{(2s)!}{2^{2s}(s!)^{2}} \right]^{2}\dfrac{l_{s}(a_{1},a_{2})}{a_{3}^{s}}={}_{2}F_{1}\left(\dfrac{1}{2},\dfrac{1}{2};1;\dfrac{\tilde f}{a_{3}} \right)\chi_{0}\\
& \tilde f^{r} \chi_{0}=s!l_{s}(a_{1},a_{2})
\end{split}\end{equation}
namely, we have reduced the series at least formally to the same hypergeometric defining the elliptic integral period. This result can be easily generalized to the case of an arbitrary product.\\

A further element of interest concerns the fact that, since, as already remarked, by replacing $\tilde f$ with $\tilde c$ the functions defining the product of Bessel and the Bessel functions are \textit{U-}equivalent, we can take advantage from the formalism to establish e.g. the \textit{n-th} derivative of the $f(x;a,b)$ functions. By noting again that it is formally written as a Gaussian, we use the following property \cite{Andrews}
\begin{equation}\begin{split}
& \tilde D_{x}^{n}e^{ax^{2}}=H_{n}(2ax,a)e^{ax^{2}}\\
& H_{n}(x,y)=n!\sum_{r=0}^{\left[\frac{n}{2}\right]}\dfrac{x^{n-2r}y^{r}}{(n-2r)!r!}
\end{split}\end{equation}
we can write the \textit{n-th} derivative of the product of two Bessel functions in terms of the two variable Hermite polynomials $H_{n}(x,y)$ as
\begin{equation}
\tilde D_{x}^{n}f(x;a,b)=  \tilde D_{x}^{n}e^{-\tilde l \left(\frac{x}{2} \right)^{2} }\Phi_{0}
=H_{n}\left(-\tilde l \frac{x}{2},-\dfrac{\tilde l}{4}\right)e^{-\tilde l \left(\frac{x}{2} \right)^{2} }\Phi_{0}=(-1)^{n}H_{n}\left(\tilde l \frac{x}{2},-\dfrac{\tilde l}{4}\right)e^{-\tilde l \left(\frac{x}{2} \right)^{2} }\Phi_{0}
\end{equation}
The use of the properties of the $-\tilde l$ operator finally yields the explicit result as
\begin{equation}\begin{split}
& \tilde D_{x}^{n}f(x;a,b)=\dfrac{(-1)^{n}}{2^{n}}n!\sum_{r=0}^{\left[\frac{n}{2}\right]}\dfrac{(-1)^{r}x^{n-2r}}{r!(n-2r)!}{}_{(n-r)}f(x;a,b)\\
& {}_{s}f(x;a,b)=\sum_{r=0}^{\infty}\dfrac{(-1)^{r}}{r!}l_{r+s}(a^2,b^2)\left(\dfrac{x}{2} \right)^{2r}
\end{split}\end{equation}
The method we have proposed here has further elements of flexibility which should be carefully examined. A more general problem, which will be just touched here and treated in a dedicated paper is the application of the formalism to the theory of multi-index Bessel functions. We remind that the Humbert functions \cite{BabDatPa} within the present formalism are defined as
\begin{equation}
I_{m_{1},m_{2}}(x)=\tilde c_{1}^{m_{1}}\tilde c_{2}^{m_{2}}e^{\tilde c_{1}\tilde c_{2}x}\varphi_{1}(0)\varphi_{2}(0)=\sum_{r=0}^{\infty}\dfrac{x^{r}}{r!(m_{1}+r)!(m_{2}+r)!}
\end{equation}
The relevant properties are easily deduced, for example we find
\begin{equation}
\tilde D_{x}I_{m_{1},m_{2}}(x)=\tilde c_{1}^{m_{1}+1}\tilde c_{2}^{m_{2}+1}e^{\tilde c_{1}\tilde c_{2}x}\varphi_{1}(0)\varphi_{2}(0)=I_{m_{1}+1,m_{2}+1}(x)
\end{equation}
Or, by applying the same integration procedure as before, we obtain
\begin{equation}\begin{split}\label{BW}
& \int_{-\infty}^{+\infty}I_{0,0}(x)e^{-\beta x^{2}}dx=\sqrt{\dfrac{\pi}{\beta}}I_{0,0}\left(\dfrac{1}{4\beta}\mid 2 \right)\\
& I_{m_{1},m_{2}}(x\mid k)=\sum_{r=0}^{\infty}\dfrac{x^{r}}{r!\Gamma (kr+1+m_{1})\Gamma (kr+1+m_{2})}
\end{split}\end{equation}
The second of eq. (\ref{BW}) is a two index Bessel-Wright equation and the Gaussian integral in the first of eq. (\ref{BW}) can be viewed as the integral transform adopted for their definition.\\

\begin{figure}[h]
 \begin{minipage}[b]{0.47\textwidth}
 \centering
 \includegraphics[width=.8\textwidth]{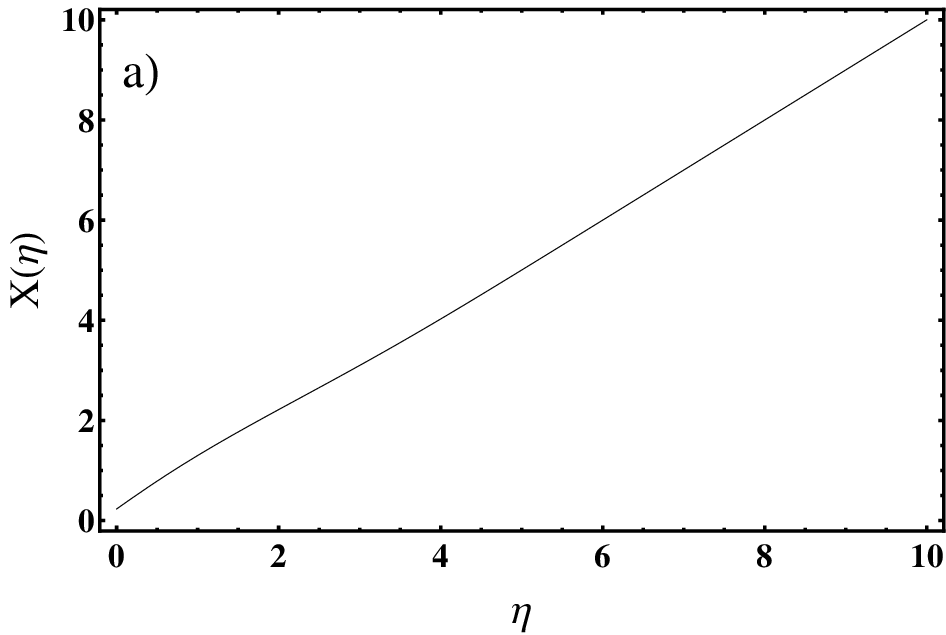}
 \end{minipage}
 \begin{minipage}[b]{0.47\textwidth}
 \centering
 \includegraphics[width=.8\textwidth]{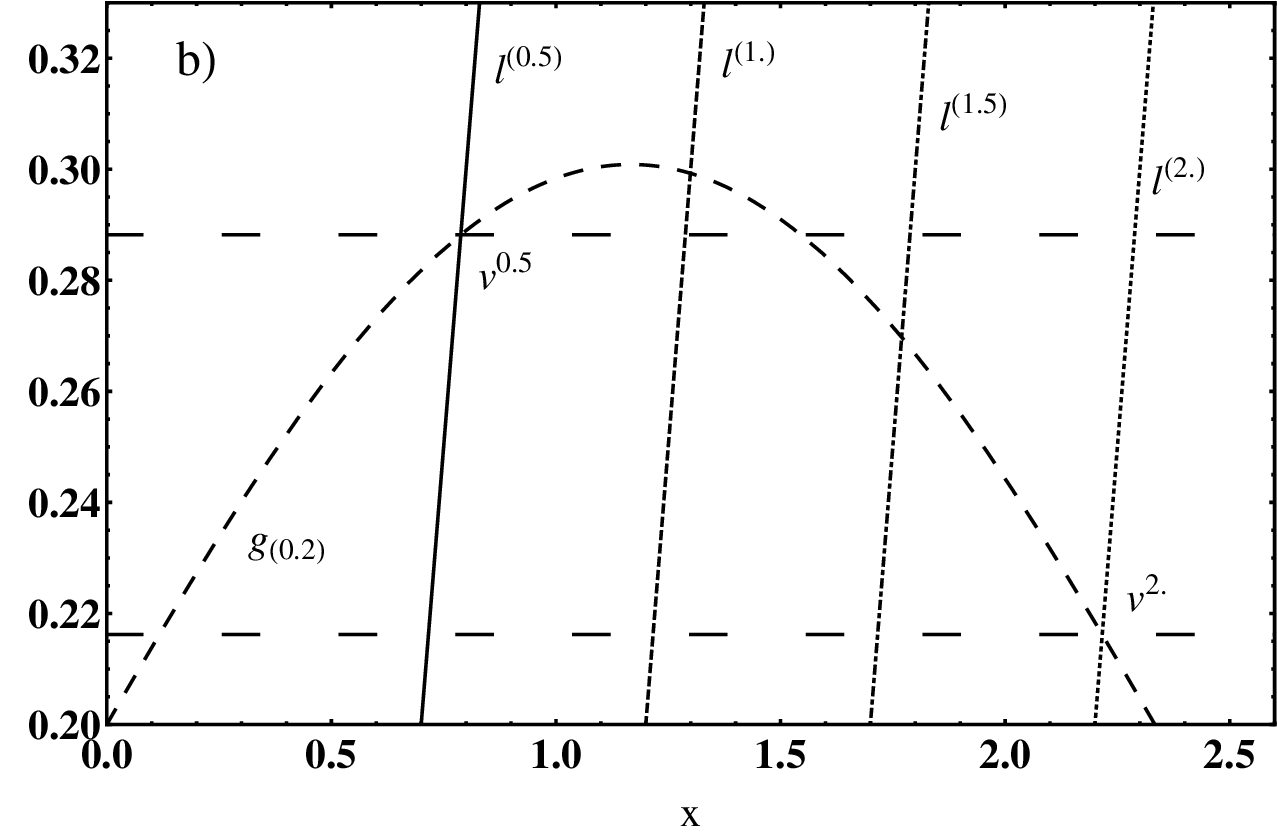}
 \end{minipage}
 \caption{Solution of the Lagrangian inversion formula for a function f given by $f(x)=e^{-a x^2 +bx}$ with $a=0.3$, $b=0.7$; a) $X(\eta)$ vs. $\eta$ for $\epsilon=0.2$; b) graphical solution of eqs.  (\ref{syst}) for different $\eta$ values at fixed $\epsilon=0.2$ and the dashed horizontal lines correspond to $\nu^{\eta}=g_{(\epsilon)}(X(\eta))$ for $\eta=0.5$ and $\eta=2$.}
 \label{fig1}
 \end{figure}

A computational application of the methods we have just discussed and more in general of the umbral procedure regarding the use of Bessel functions is an extension of the Lagrange expansion method \cite{Goursat} for the solution of non- linear algebraic equation involving special functions.\\
To this aim we remind that if $f(x)$ is any continuous infinitely differential function in a point a, then the solution of the equation
\begin{equation}
x=\eta+\epsilon f(x)
\end{equation}
can be written as
\begin{equation}
x=\eta+\sum_{n=1}^{\infty} \frac{\epsilon^n}{n!}(\widehat{D}_x^{n-1}\left[f(x)\right]^n)|_{x=a}
\end{equation}
Assuming e.g. $f(x)=e^{-ax^2+bx}$ we obtain the solution of our problem in the form
\begin{equation}
X(\eta)=\eta+\sum_{n=1}^{\infty} \frac{\epsilon^n}{n!}H_{n-1}(-2n \eta+b n,-a n) e^{-a n \eta^2+bn \eta}
\label{eq:num3}
\end{equation}
In Fig.\ref{fig1}a we have accordingly reported $X(\eta)$ vs. $\eta$ for some values of the parameters, while in Fig.\ref{fig1}b we have proven the correctness of our solution by checking the coincidence of the interception between the curves
\begin{equation}
\left\{ \begin{array}{cc}
l^{(\eta)}(x)=x-\eta  \\ \\
g_{(\epsilon)}(x)=\epsilon f(x)
\end{array}
\right.
\label{syst}
\end{equation}
and the numerical solutions reported in Fig.\ref{fig1}a .

The range of parameters have been chosen to ensure the convergence of the  series on the rhs of eq.\ref{eq:num3}. Without entering the details concerning the range of validity of the solution \ref{eq:num3} we consider its extension to the case of Bessel functions. \\
We consider therefore the equation
\begin{equation}
x=\eta+\epsilon J_0(x)
\end{equation}
an d the relevant solution in the form

\begin{equation}
x=\eta+\sum_{n=1}^{\infty} \frac{\epsilon^n}{n!}(\widehat{D}_x^{n-1}\left[J_0(x)\right]^n)|_{x=a}
\end{equation}
The use of generalization of the Leibniz rule on repeated product of derivatives yields

\begin{equation}
\partial_x^m(f_1...f_m)=\sum_{k_1+...+k_m=n}^m \left( \begin{array}{c} m  \\ k_1,...,k_m \end{array} \right) \prod_{r=1}^{m}\partial_x^{k_r} f_r
\end{equation}

Where the multifactorial is defined as

\begin{equation}
\left( \begin{array}{c} n  \\ k_1,...,k_m \end{array} \right)=\frac{n!}{k_1!...k_m!}
\end{equation}
The use of our umbral notation yields

\begin{equation}
x=\eta+\sum_{n=1}^{\infty} \frac{\epsilon^n}{n!} \sum_{k_1+...+k_m=n}^{n-1} \left( \begin{array}{c} n-1  \\ k_1,...,k_m \end{array} \right)\prod_{r=1}^{m} (-1)^r H_r\left( \frac{1}{2} \widetilde{c}_r ,x,-\frac{1}{2} \widetilde{c}_r \right) e^{ \widetilde{c}_r \left(\frac{x}{2}\right)^2 } \widetilde{\varphi}_r
\end{equation}

In this paper we have shown that a formalism of umbral nature can be exploited to simplify in a significant way the technicalities underlying the theory of Bessel functions and of their manipulations leading to combinations or to the introduction of new forms. In forthcoming investigations we will further extend the method.

\end{document}